\documentclass[manuscript,sort&compress]{elsarticle}
\usepackage[dvips,a4paper,width=18cm,height=26cm]{geometry}
\usepackage{amsmath}
\usepackage{amssymb}
\usepackage{graphicx}

\title{Optimal Learning Dynamics of Multi Agents
in Restless Multiarmed Bandit Game}

\author{Kazuaki Nakayama\corref{cor1}}
\ead{nakayama@math.shinshu-u.ac.jp}
\address{Department of Mathematics, Faculty of Science, Shinshu University, Asahi 3-1-1, Matsumoto, Nagano 390-8621, Japan}

\author{Ryuzo Nakamura}
\address{Department of Physics, Faculty of Science, Kitasato University,
Kitasato 1-15-1, Sagamihara, Kanagawa 252-0373, Japan}

\author{Masato Hisakado}
\address{Nomura Holdings Inc.,
Otemachi 2-2-2, Chiyoda-ku, Tokyo 100-8130, Japan}

\author{Shintaro Mori}
\address{Department of Mathematics and Physics,
Faculty of Science and Technology, Hirosaki University,
Bunkyo-cho 3, Hirosaki, Aomori 036-8561, Japan}

\cortext[cor1]{Corresponding author}

\begin{document}

\begin{abstract}
  Social learning is learning through the observation of or
  interaction with other 
individuals; it is critical in the understanding of
the collective behaviors of humans in social physics.
We study the learning process of agents in a restless multiarmed bandit (rMAB).
The binary payoff of each arm changes randomly and agents maximize their
payoffs by exploiting an arm with payoff 1,
searching the arm at random (individual learning), or copying an arm
exploited by other agents (social learning).
The system has Pareto and Nash equilibria in the mixed strategy space
of social and individual learning.
We study several models in which agents maximize their expected payoffs in
the strategy space, and demonstrate analytically and numerically that
the system converges to the equilibria.
We also conducted an experiment and investigated
whether human participants adopt the optimal strategy.
In this experiment, three participants play the game.
If the reward of each group is proportional to the sum of the payoffs,
the median of the social learning rate almost coincides with that
of the Pareto equilibrium.
\end{abstract}


\maketitle

\section{Introduction}

A multiagent system is an active research field and it is applicable
 to many academic disciplines. In physics, it has been employed to
study the collective behaviors of humans in econophysics\cite{MantegnaStanley,Ormerod}
and socio-physics\cite{Galam2008,CastellanoFortunatoLoreto2009,Pentland}.
If the system exhibits intriguing macroscopic behaviors,
it is challenging to solve the system theoretically.
For example, the voter model abstracts the influence process
among humans and exhibits rich mathematical structures, as a
consensus formation\cite{Mobilia2003,MobiliaPetersenRedner2007,%
SoodRedner2005,SanoHisakadoMori}.
In the studies, the model parameters are typically constant and
the agents do not learn from their experiences.
If the agents learn from their experience,
the problem becomes an reinforcement learning multiagent system.
The model parameters change as the agents learn;
therefore, it becomes more difficult to solve the system than the nonlearning agent system.

Social learning is learning through the observation of or interaction with other
individuals\cite{Laland2004,KendalGiraldeauLaland2009,Rendell,Kameda,Toyokawa}.
As it causes the tendency to follow others' behaviors, the system of social
learning agents becomes a strongly correlated system.
As a toy model, we have proposed a multiagent system\cite{EU11518}
in a restless multiarmed bandit (rMAB).
A restless multiarmed bandit\cite{Rendell} is a slot machine with multiple arms.
The term ``restless'' implies that the payoffs change randomly.
We call an arm with a high payoff a good arm.
Each agent maximizes his payoffs by exploiting an arm,
searching for a good arm (individual learning), and copying an arm
exploited by other agents (social learning).
The system exhibits a phase transition at a critical value of the social learning
probability\cite{EU11518}.
If the social learning probability is below the threshold value, the variance
of the number of agents who
have found the good arm is proportional to the number of agents. If it exceeds the
threshold value,
the variance becomes proportional to the power of the number of agents with
a critical exponent larger than 1.
The system shows an oscillatory behavior between the state where almost all
agents know the good arms and
the state where none of the agents know them. Further, the optimal value of the social learning
probability is studied.
In \cite{NakayamaHisakadoMori2017}, we studied an rMAB with only one good
arm and demonstrated that the system yields the unique Nash equilibrium and 
Pareto optimal states, thus solving
the famous Rogers paradox\cite{Rogers,Boyd1995,Enquist,Rendell2010}.
The question that arises is how the agents adopted the optimal strategies.
If the agents can change their social learning probability, the optimization
problem can be formulated as a reinforcement learning problem.
It is a difficult problem and an efficient approach does not exist.
In this study, we investigate the type of strategies that agents adopt
when they attempt to maximize their fitness.
Hence, we first analytically and numerically study the best response
dynamics and its natural variants, that is, whether they
reach one of the natural equilibria, the Nash equilibrium, and the Pareto
optimal states.
Next, we replace the agents with humans and perform an experiment to
examine whether humans tend to employ the optimal strategy\cite{Yoshida}.
In this experiment, three human players play the rMAB game.

This manuscript is organized as follows. In the next two sections, we state the
definition of our model and present the
necessary results from our previous work\cite{NakayamaHisakadoMori2017}.
In section 4, we study the best response dynamics. We demonstrate that they
converge to the Nash equilibrium or the Pareto optimal state. In section 5, we
elaborate the experimental setup and present the results.
The last section presents the concluding remarks.

\section{Model}
\label{sec:Model}

\begin{figure}[ht]
  \centering
  \includegraphics[width=13cm]{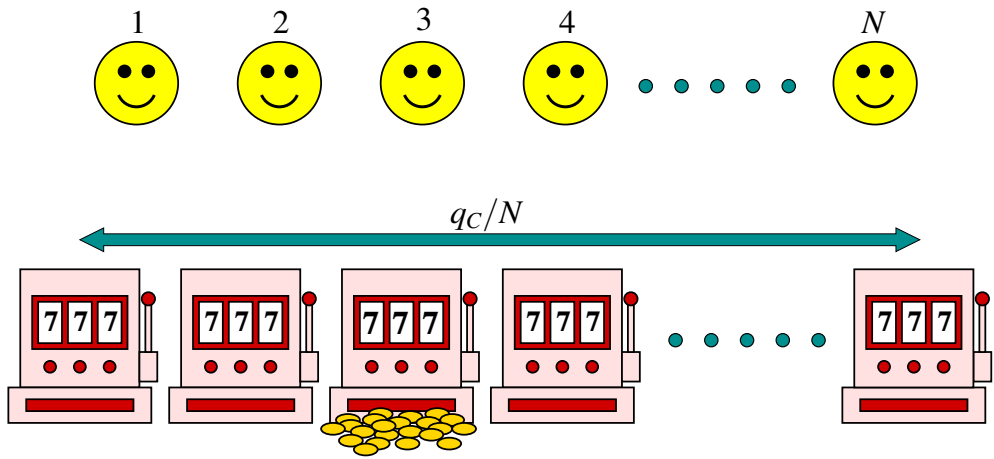}
  \caption{Bandits and agents.}
  \label{fig:bandits-agents}
\end{figure}

The rMAB contains only one good arm and infinite number of bad arms.
There are $N$ agents labeled by $n=1,\cdots,N$ (see fig.~\ref{fig:bandits-agents}).
The system evolves in time as follows.

In each turn, an agent (e.g., agent $n$) is chosen randomly.
He exploits his arm and obtains payoff $1$ if he knows a good arm.
If he does not know a good arm, he searches for it by a random search
(individual learning) with probability $1-r_{n}$,
or copies the information of other agents' good arms (social learning)
with probability $r_{n}$.
The random search always succeeds with probability $q_{I}$.
Meanwhile, the copy process succeeds with probability $q_{O}$
if and only if at least one agent knows a good arm.

Subsequently, with probability $q_{C}/N$, the good arm changes to a bad arm
and another good arm appears (environmental change).
When an environmental change occurs, the agents who know the good arm
are forced to forget it and know a bad one.

This completes the turn.
The system will proceed to the next turn.

We call $N$ consecutive turns a step.
It is expected that each agent performs one action per step.
The probability that the environment does not change during one step
is $(1-q_{C}/N)^{N}$ which is close to $e^{-q_{C}}$ when $N$ is large.
Therefore, this probability must not be small
for the copy process to be meaningful.

\begin{figure}[ht]
  \centering
  \includegraphics[width=10cm]{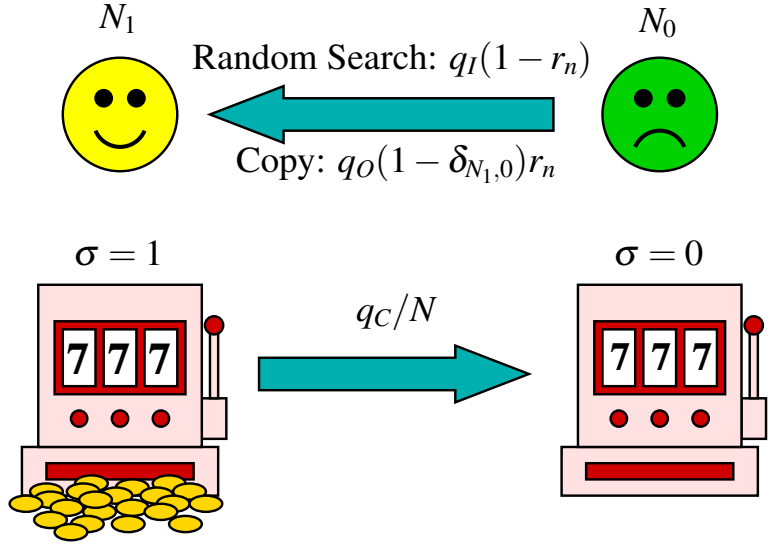}
  \caption{Time evolution of the state of the system.  $N_{0}$ and $N_{1}$
  are the number of agents who know/do not know a good arm, respectively.
  The entity $\sigma$ is defined in \eqref{eq:sigma}.}
  \label{fig:time-evolution}
\end{figure}

Figure \ref{fig:time-evolution} shows the schematics of 
the time evolution of the system.

\newpage

\paragraph{Mathematical Formulation}

Next, we formulate a stochastic model.
We introduce the random variable $\sigma_{n}$ defined by
\begin{equation}
  \label{eq:sigma}
  \sigma_{n} =
  \begin{cases}
    1, & \text{if agent $n$ knows a good arm},
    \\
    0, & \text{if agent $n$ does not know a good arm}.
  \end{cases}
\end{equation}
We regard $\sigma_{n}$ as agent $n$'s knowledge of a good arm.
For each turn $t$, we have a joint probability function
$P(\sigma_{1},\cdots,\sigma_{N}|t)$. For simplicity we use vector notation: $P(\vec{\sigma}|t)$,
$\vec{\sigma}=(\sigma_{1},\cdots,\sigma_{N})$.
Time evolution is described by the Chapman--Kolmogorov equation:
\begin{equation}
  \label{eq:Chapman-Kolmogorov}
  P(\vec{\sigma}^{\prime}|t+1) = \sum_{\vec{\sigma}}T(\vec{\sigma}^{\prime}|\vec{\sigma})P(\vec{\sigma}|t),
\end{equation}
where $T(\vec{\sigma}^{\prime}|\vec{\sigma})$ is
the stochastic matrix (transfer matrix) of the system.
The matrix is expressed as the product of the agent action part and
the environmental change part:
\begin{equation}
  \label{eq:TransferMatrix}
  T(\vec{\sigma}^{\prime}|\vec{\sigma}) = \sum_{\vec{\sigma}^{\prime\prime}}
  T^{C}(\vec{\sigma}^{\prime}|\vec{\sigma}^{\prime\prime})T^{A}(\vec{\sigma}^{\prime\prime}|\vec{\sigma}).
\end{equation}
The environmental change part, $T^{C}(\vec{\sigma}^{\prime}|\vec{\sigma})$,
is obtained as follows.
When no environmental change occurs, each agent stores his knowledge
on the good arm.  If an environmental change occurs, all the agents
lose their knowledge. Thus, we have
\begin{equation}
  \label{eq:CTransferMatrix}
  T^{C}(\vec{\sigma}^{\prime}|\vec{\sigma})
  = \left(1-\frac{q_{C}}{N}\right)\prod_{n=1}^{N}\delta_{\sigma_{n}^{\prime}\sigma_{n}}
  + \frac{q_{C}}{N}\prod_{n=1}^{N}\delta_{\sigma_{n}^{\prime}0},
\end{equation}
where $\delta_{\sigma_{n}\sigma_{m}}$ is the Kronecker delta.
Next, we examine the agent action part, $T^{A}(\vec{\sigma}^{\prime}|\vec{\sigma})$.
The probability that agent $n$ who does not know the good arm
is chosen and finds a good arm is
\begin{align}
  \label{eq:p_n}
  p_{n}(\vec{\sigma}) &= \frac{\delta_{\sigma_{n}0}}{N}\{r_{n}(1-\delta_{N_{1}0})q_{O}
  + (1-r_{n})q_{I}\},
  \\
  \label{eq:N1}
  N_{1} &= \sum_{n=1}^{N}\sigma_{n}.
\end{align}
The matrix $T^{A}(\vec{\sigma}^{\prime}|\vec{\sigma})$ is the sum of the
probability that no agent changes his knowledge
and the probability that one of the agents changes his knowledge.
Thus, we have
\begin{align}
  \label{eq:ATransferMatrix}
  T^{A}(\vec{\sigma}^{\prime}|\vec{\sigma})
  &= p_{NC}(\vec{\sigma})\prod_{n=1}^{N}\delta_{\sigma_{n}^{\prime}\sigma_{n}}
  + \sum_{n=1}^{N}p_{n}(\vec{\sigma})\delta_{\sigma_{n}^{\prime},\sigma_{n}+1}
  \cdot\prod_{\ell\neq n}\delta_{\sigma_{\ell}^{\prime}\sigma_{\ell}},
  \\
  \label{eq:p_NC}
  p_{NC}(\vec{\sigma}) &= 1 - \sum_{n=1}^{N}p_{n}(\vec{\sigma}).
\end{align}
This completes the formulation of our stochastic model.

We make a short comment on the assumption of social learning.
In our model, the copy process is assumed to succeed with probability
$q_{O}(1-\delta_{N_{1}0})$.  As already mentioned, this means that
social learning succeeds with probability $q_{O}$ if and only if at least
one agent knows the correct answer.
However, what we want to emphasize is that the copy process always fails if
no agent knows the answer.  Our intention is to rule out copying of incorrect information\cite{Rendell,Enquist}.
For this purpose, it is possible to replace the factor ``$1-\delta_{N_{1}0}$'',
for example, with $N_{1}/N$, the proportion of agents who know the correct answer.
The new model is a natural replacement for the original one.
Unfortunately, it can not be solved analytically.
A numerical simulation indicates that the new model seems to behave similarly to
the original one with a smaller $q_{O}$.

\paragraph{Asymptotic Property of $P(\vec{\sigma}|t)$}

It is proven that the matrix $T(\vec{\sigma}^{\prime}|\vec{\sigma})$
is irreducible and primitive\cite{CarlDMeyer} if and only if
\begin{equation}
  \label{eq:assumption0}
  q_{C}>0,\quad
  q_{I}>0,\quad
  \forall r_{n}<1.
\end{equation}
Under this assumption, the unique Perron vector $P(\vec{\sigma})$ exists.
The Perron vector has the remarkable feature that
it is the long-time limit
of an arbitrary solution $P(\vec{\sigma}|t)$ of the Chapman--Kolmogorov equation,
that is, $\displaystyle{\lim_{t\rightarrow\infty}P(\vec{\sigma}|t)=P(\vec{\sigma})}$.
We say that the system is in the steady state when the probability function
is $P(\vec{\sigma})$.

The irreducibility of the matrix $T(\vec{\sigma}^{\prime}|\vec{\sigma})$ presents a physical
meaning. Let $\vec{\sigma}$ be a knowledge vector at an instant turn $t$.
Then, for an arbitrary knowledge vector $\vec{\sigma}^{\prime}$,
the probability that the change $\vec{\sigma}\rightarrow\vec{\sigma}^{\prime}$
occurs after a few turns is positive (this is merely the definition
of irreducibility).
Indeed, $q_{C}=0$ implies that the knowledge on a good arm will never be lost
and the dynamics of the system may become trivial.
In other words, a transition of type $\vec{\sigma}\neq\vec{0} \rightarrow\vec{0}$ does not exist.
When $q_{I}(1-r_{n})=0$, the transition $\vec{0}\rightarrow
(0,\cdots,0,\overset{n}{\check{1}},0,\cdots,0)$
is impossible.
Thus, no agent will obtain a correct information on a good arm
when the initial condition is given by $\vec{\sigma}=\vec{0}$.
Conversely, if \eqref{eq:assumption0} is satisfied,
an arbitrary transition $\vec{\sigma}\rightarrow\vec{\sigma}^{\prime}$
is possible through $\vec{0}$. This proves the irreducibility of the probability matrix.

When the matrix $T$ is irreducible, it is also primitive because
$\operatorname{tr}T \ge T(\vec{0}|\vec{0})>0$.
Primitivity ensures that the long-time limit of the probability function
$\displaystyle{\lim_{t\rightarrow\infty}P(\vec{\sigma}|t)}$
coincides with the Perron vector $P(\vec{\sigma})$
regardless of the initial probability function $P(\vec{\sigma}|0)$.

In the following, we assume that the condition
\eqref{eq:assumption0} holds,
in addition to the inequality
\begin{equation}
  \label{eq:assumption1}
  q_{O} > 0
\end{equation}
for nontriviality.

\section{Nash Equilibrium and Pareto Optimality}
\label{sec:NashPareto}

\paragraph{Fitness Function}

We summarize the necessary results
from our previous work\cite{NakayamaHisakadoMori2017}.
The expected payoff for each agent in the steady state is defined by
\begin{equation}
  \label{eq:profit}
  w_{n} = E[\sigma_{n}] = \sum_{\vec{\sigma}}P(\vec{\sigma})\sigma_{n},\quad
  n=1,\cdots,N.
\end{equation}
We call $w_{n}$ the fitness of agent $n$.
It is shown
that a function $w(r,\overline{r})$ exists such that
$w_{n}$ is expressed as
\begin{align}
  w_{n} &= w(r_{n},\overline{r}_{n}),
  \quad
  \overline{r}_{n} = \frac{1}{N-1}\sum_{k\neq n}r_{k}.
  \label{eq:profit_n}
\end{align}
We call $w(r,\overline{r})$ the fitness function, which is expressed explicitly by
\begin{gather}
  \label{eq:ProfitFunction}
  w(r,\overline{r})
  = \frac{1}{a+q_{I}+(q_{O}-q_{I})r}
  \left\{q_{I}+(q_{O}-q_{I})r
    - \frac{aq_{O}r}{a+\kappa}
  \right\},
  \\
  \label{eq:kappa_a}
  \kappa = (N-1)q_{I}(1-\overline{r})+q_{I}(1-r),
  \quad
  a = \frac{q_{C}}{1-q_{C}/N}.
\end{gather}

\paragraph{Nash Equilibrium}

The strategy space of the agents is an $N$-dimensional unit cube,
\begin{equation}
  \label{eq:strategy-space}
  J = \{\vec{r}=(r_{1},\cdots,r_{N})\,|\,0\le r_{i}\le 1\}.
\end{equation}
It is noteworthy that the limit $r_{i}\rightarrow1$ is not excluded.
In this space, the unique Nash equilibrium point $\vec{r}_{\text{Nash}}$ exists.
It exhibits the following properties:
(i) it is symmetric, $\vec{r}_{\text{Nash}}=(r_{\text{Nash}},\cdots,r_{\text{Nash}})$,
(ii) $r_{\text{Nash}}<1$,
(iii) $\vec{r}_{\text{Nash}}$ is strict,
\begin{equation}
  w(r_{0},r_{\text{Nash}})
  < w(r_{\text{Nash}},r_{\text{Nash}}),\quad
  {}^{\forall}r_{0}\neq r_{\text{Nash}},
  \label{eq:ESS1}
\end{equation}
and (iv) $r_{\text{Nash}}\rightarrow0$ as $\{N(q_{O}-q_{I})-(a+q_{O})\}\rightarrow+0$.
The specific form of $r_{\text{Nash}}$ is as follows:
\begin{equation}
  \label{eq:NashEquilibrium}
  r_{\text{Nash}}
  =
  \begin{cases}
    1 - \eta, & N(q_{O}-q_{I}) > a+q_{O},
    \\
    0, & N(q_{O}-q_{I}) \le a+q_{O},
  \end{cases}
\end{equation}
where
\begin{align*}
  \eta &= \frac{2(a+q_{O})^{2}}{(q_{O}-q_{I}N)(a+q_{O}) + (aN+q_{O})(q_{O}-q_{I})+\sqrt{D_{1}}},
  \\
  D_{1}
  &= \{(q_{O}-q_{I}N)(a+q_{O}) - (aN+q_{O})(q_{O}-q_{I})\}^{2}
  + 4(N-1)(q_{O}-q_{I})(a+q_{O})q_{O}(a+Nq_{I}).
\end{align*}

Because $\vec{r}_{\text{Nash}}$ is strict, it is 
an evolutionarily stable strategy (ESS)\cite{Maynard-Smith-Price}.
Moreover, this is an ESS based on Thomas\cite{Thomas1}.
Indeed, the following inequality is true:
\begin{equation}
  w(r_{\text{Nash}},r_{0}) > w(r_{0},r_{0}),\quad
  {}^{\forall}r_{0}\neq r_{\text{Nash}}.
  \label{eq:ESS2}
\end{equation}

\paragraph{Pareto Optimality}

The concept of Pareto optimality is defined in the context of resource allocation.
We have regarded the fitness $w_{n}=w(r_{n},\overline{r}_{n})$ as the utility of
agent $n$. Let us also consider this entity as the amount of resources
acquired by agent $n$.
At this time, the maximum point of the total utility function,
\begin{equation}
  \label{eq:total_fitness_function}
  I(\vec{r}) = \sum_{n=1}^{N}w_{n}(r_{n},\overline{r}_{n}),
\end{equation}
is suitable as a Pareto optimal point.
It is shown in the Appendix that
only one maximum point exists, which we call $\vec{r}_{\text{Pareto}}$.
It is strictly maximal and symmetric:
$\vec{r}_{\text{Pareto}}=(r_{\text{Pareto}},\cdots,r_{\text{Pareto}})$.

It is clear that $r_{\text{Pareto}}$ is 
the only maximum point of $w(r,r)$.
Our result is that
\begin{equation}
  \label{eq:ParetoOptimal}
  r_{\text{Pareto}}
  =
  \begin{cases}
    \dfrac{(a+q_{I}N)X - (a+q_{I})Y}{q_{I}NX + (q_{O}-q_{I})Y},
    & N(q_{O}-q_{I}) > a+q_{O},
    \\
    0, & N(q_{O}-q_{I}) \le a+q_{O},
  \end{cases}
\end{equation}
where
\begin{align*}
  X &= \sqrt{(N-1)(a+q_{O})(q_{O}-q_{I})},
  \\
  Y &= \sqrt{Nq_{O}(a+Nq_{I})}.
\end{align*}
It is not difficult to verify that
(i) $r_{\text{Pareto}}<1$, and (ii) $r_{\text{Pareto}}\rightarrow0$
as $\{N(q_{O}-q_{I})-(a+q_{O})\}\rightarrow+0$.

\paragraph{Comparison of Fitness}

We consider three types of fitness per agent:
the fitness of individual learners,
that in the Nash equilibrium state,
and that in the Pareto optimal state.
They are defined as follows, respectively:
\begin{equation}
  \label{eq:w_I-w_Nash-w_Pareto}
  w_{I} = w(0,\overline{r}),\quad
  w_{\text{Nash}} = w(r_{\text{Nash}}, r_{\text{Nash}}),\quad
  w_{\text{Pareto}} = w(r_{\text{Pareto}}, r_{\text{Pareto}}).
\end{equation}
It is proven that the inequality,
$w_{I} \le w_{\text{Nash}} \le w_{\text{Pareto}}$,
is true.
The equality holds if and only if
$N(q_{O}-q_{I})\le a+q_{O}$.

\section{Optimal Strategies of Agents}
\label{sec:OptimalStrategy}

Nash equilibrium and Pareto optimality are concepts in equilibrium.
Thus, it is important to demonstrate that some learning processes
converge to the Nash equilibrium and Pareto optimal states.
A standard model is the best response dynamics, which is expected to
reach the Nash equilibrium state.
In this section, we study the best response dynamics and its variants.
Best response dynamics seems to have been appeared by simplifying fictitious
play\cite{Hopkins1999,GilboaMatsui1991,FudenbergLevine}.
One of the purpose of both
learning models is to explain Nash equilibria by each agent's local search method of strategy\cite{FeldmanSnappirTamir}.

\subsection{Best Response Dynamics}

Several versions of the best response dynamical system exist.
Herein, we consider the continuous-time best response dynamical system.

\begin{figure}[h]
  \centering
  \includegraphics[width=10cm]{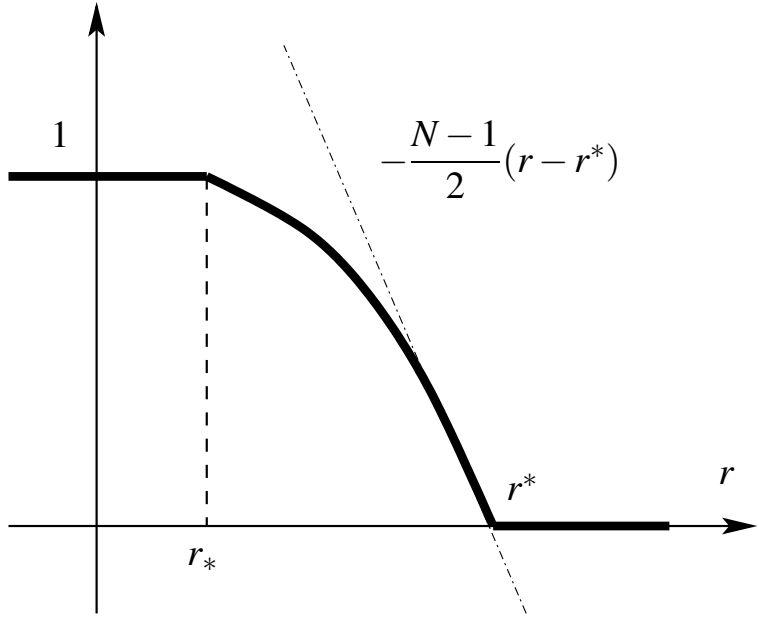}
  \caption{Plot of the best response function $\beta_{N}(r)$.
  The entities $r^{\ast}$ and $r_{\ast}$ are expressed as $r^{\ast}=1-\frac{a+q_{I}}{(N-1)(q_{O}-q_{I})}$ and
  $r_{\ast}=1-\frac{q_{I}(a+q_{O})-(q_{O}-q_{I})a+\sqrt{D_{2}}}{2(N-1)q_{I}(q_{O}-q_{I})}$, where
  $D_{2}=\{q_{I}(a+q_{O})-(q_{O}-q_{I})a\}^{2}
  + 4q_{I}(q_{O}-q_{I})(a+q_{O})^{2}$.}
  \label{fig:beta}
\end{figure}

First, we introduce the best response function,
\begin{equation}
  \label{eq:BestResponseFunction}
  \beta_{N}(\overline{r}) = \mathop{\mathrm{argmax}}\limits_{0\le r\le 1}w(r,\overline{r}).
\end{equation}
The best response function of agent $n$ is expressed as $\beta_{N}(\overline{r}_{n})$.
See eqs.~\eqref{eq:profit_n} and \eqref{eq:ProfitFunction}.
The properties of the function $\beta_{N}(r)$ are summarized in fig.~\ref{fig:beta}.

The continuous-time best response dynamics is a learning process
defined by the following dynamical system:
\begin{equation}
  \frac{dr_{n}}{dt}(t) = \beta_{N}(\overline{r}_{n}(t)) - r_{n}(t),
  \quad
  n=1,\cdots,N.
  \label{eq:continuous-brd}
\end{equation}
Because the range of the function $\beta_{N}(r)$ is included in the
closed interval $[0,1]$, this is a well-defined system of
ordinary differential equations on the strategy space $J$.
It is known that not all dynamical systems of this type converge to
a Nash equilibrium point\cite{Roughgarden}.
Nonetheless, we shall demonstrate that any solution to equation
\eqref{eq:continuous-brd} converges to the unique Nash
equilibrium point $\vec{r}_{\text{Nash}}$.

The following relation
derived from fig.~\ref{fig:beta}
is noteworthy:
\begin{equation}
  \beta_{N}(r) - \beta_{N}(r_{\text{Nash}}) = -(N-1)\gamma(r)(r-r_{\text{Nash}}),\quad
  0 \le \gamma(r) \le \frac{1}{2}.
  \label{eq:beta-estimation}
\end{equation}
We write $\gamma(\overline{r}_{n}(t))$ as $\gamma_{n}(t)$ for brevity.
Let $x_{n}(t) = r_{n}(t) - r_{\text{Nash}}$.  
Because $r_{\text{Nash}}$ is the fixed point of $\beta_{N}(r)$, 
the next equation is derived,
\begin{equation}
  \beta_{N}(\overline{r}_{n}(t)) - r_{n}(t)
  = -\gamma_{n}(t)\sum_{k\neq n}x_{k}(t) - x_{n}(t).
  \label{eq:beta-gamma}
\end{equation}
We introduce the positive definite quadratic form:
\begin{equation}
  V(\vec{x}) = \frac{1}{2}\sum_{n=1}^{N}x_{n}^{2}
  + \frac{1}{2}\left(\sum_{n=1}^{N}x_{n}\right)^{2}.
  \label{eq:LyapunovFunction}
\end{equation}
The function $V(\vec{x})$ contains a unique minimum at $\vec{x}=\vec{0}$.
We differentiate $V(\vec{x})$ along a solution of 
\eqref{eq:continuous-brd},
\begin{align*}
  \frac{d}{dt}V(\vec{x}(t))
  &= \sum_{n=1}^{N}\frac{dx_{n}}{dt}(t)\left\{
    x_{n}(t) + \sum_{k=1}^{n}x_{k}(t)\right\}
  \\
  &= -\sum_{n=1}^{N}(1-\gamma_{n}(t))x_{n}(t)^{2}
  - \left\{1+\sum_{n=1}^{N}\gamma_{n}(t)\right\}
  \left(\sum_{k=1}^{n}x_{k}(t)\right)^{2}
  \\
  &\le -\frac{1}{2}\sum_{n=1}^{N}x_{n}(t)^{2}
  - \frac{1}{2}\left(\sum_{n=1}^{N}x_{n}(t)\right)^{2}
  \\
  &= -V(\vec{x}(t)).
\end{align*}
This yields
\begin{equation*}
  0 \le V(\vec{x}(t)) \le V(\vec{x}(0))e^{-t} \rightarrow 0
  \quad(t\rightarrow\infty).
\end{equation*}
Thus, it is proven that an arbitrary solution $\vec{r}(t)$ of 
\eqref{eq:continuous-brd} converges to $\vec{r}_{\text{Nash}}$.

It is noteworthy that the discrete best response dynamics,
\begin{equation}
  \label{eq:discrete-brd}
  r_{n}(t+1) = r_{n}(t) + \alpha\{\beta_{N}(\overline{r}_{n}(t)) - r_{n}(t)\},\quad
  n=1,\cdots,N,
\end{equation}
also converges to the Nash equilibrium irrespective of
the initial condition $\vec{r}(0)\in J$ if the learning rate $\alpha$
is sufficiently small\footnote{A sufficient
condition is that $\alpha\le\frac{2}{4+N\{N+2+(N+3)^{2}\}}$.}.
This is evident because \eqref{eq:continuous-brd} is a continuous
limit $\alpha\rightarrow+0$ with substitution $r_{n}(t+1)\leftarrow
r_{n}(t+\alpha)$
in the left-hand side\  of \eqref{eq:discrete-brd}.

Further, in eq.~\eqref{eq:discrete-brd},
the strategies of all agents are updated simultaneously.
Meanwhile, a model exists in which
only the randomly selected agent's strategy is updated.
The latter is represented by replacing the learning rate $\alpha$
with $\alpha/N$ in eq.~\eqref{eq:discrete-brd}.

\subsection{Derivative Best Response Dynamics}

The best response dynamics is a natural and straightforward learning
procedure that can lead the system to the Nash equilibrium.
However, when no agent can obtain other agents' strategies,
it is impossible for agent $n$ to know the value, $\beta_{N}(\overline{r}_{n}(t))$,
of the best response function.

\begin{figure}[ht]
  \centering
  \includegraphics[width=8cm]{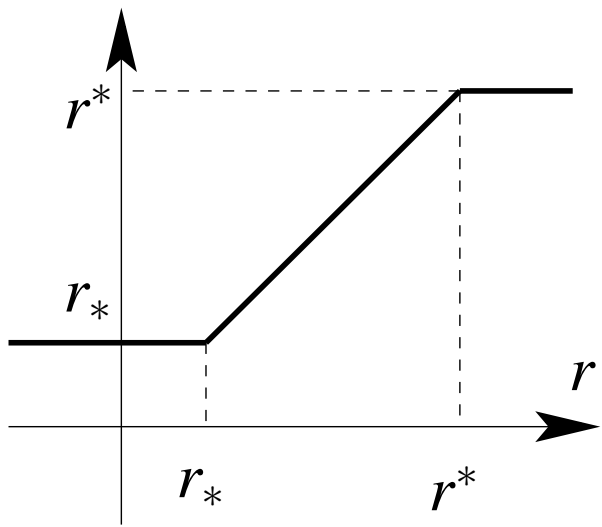}
  \caption{Plot of the function $[r]$.
  The entities $r^{\ast}$ and $r_{\ast}$ are defined in
  fig.~\ref{fig:beta}.}
  \label{fig:urfunc}
\end{figure}

Therefore, as a ``realistic'' model, we introduce the following system:
\begin{equation}
  \frac{dr_{n}}{dt}(t) = \frac{\partial w}{\partial r}(r_{n}(t),[\overline{r}_{n}(t)]),
  \quad n=1,\cdots,N,
  \label{eq:derivative-brd}
\end{equation}
where $[r]\equiv\max(r_{\ast},\min(r^{\ast},r))$ is the truncation of $r$ to the closed
interval $[r_{\ast},r^{\ast}]$ (see fig.~\ref{fig:urfunc}).
Because $w(r_{n},\overline{r}_{n})$ is the 
expected
income of agent $n$, he could
estimate the differential coefficient,
$\dfrac{\partial w}{\partial r}(r_{n}(t),[\overline{r}_{n}(t)])$.

In reinforcement learning, it is natural to replace the estimate by a numerical
derivative.
Agent $n$ estimates $w(r_n,\overline{r}_n)$ by the time average of his income
$\sigma_n$ for $r_n,\overline{r}_n$.
 He changes $r_n$ to $r_n+\Delta r_n$ and
estimates the time average. We divide the difference in the time averages
by $\Delta r_n$ and obtain the derivative.

The fitness function $w(r,\overline{r})$ contains a unique maximal point
as a function of the first variable $r$. This point is a strictly
decreasing function of the second variable $\overline{r}$.
Let $\overline{\beta}_{N}(\overline{r})$ be the function thereof.
Subsequently, the following relation is true:
\begin{equation*}
  \beta_{N}(\overline{r}) = \overline{\beta}_{N}([\overline{r}]) = \max(0,\min(1,\overline{\beta}_{N}(\overline{r}))).
\end{equation*}

As a consequence, the crucial condition
\begin{equation*}
  0 \le \frac{\partial w}{\partial r}(0,[\overline{r}_{n}(t)]),\quad
  \text{and}\quad
  \frac{\partial w}{\partial r}(1,[\overline{r}_{n}(t)])\le0,\quad
  n=1,\cdots,N,
\end{equation*}
holds.  This ensures that any solution $\vec{r}(t)$ of the system
\eqref{eq:derivative-brd} remains in the domain $J$.
Therefore, the introduction of the function $[r]$ renders
\eqref{eq:derivative-brd} well-defined.

\begin{figure}[h]
  \centering
  \includegraphics[width=8cm]{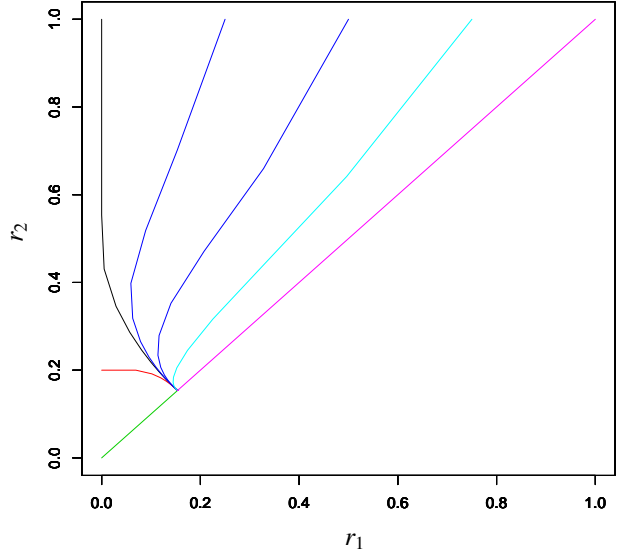}
  \caption{Numerical integration of the derivative best response dynamical system
  for various initial points.
  Parameters: $q_{C}=0.1$, $q_{I}=0.2$, $q_{O}=0.8$, $N=2$.
  Our analytical result is that $r_{\text{Nash}}=0.1537...$.}
  \label{fig:integral-curve}
\end{figure}
\begin{figure}[h]
  \centering
  \includegraphics[width=12cm,bb=0 0 450 300]{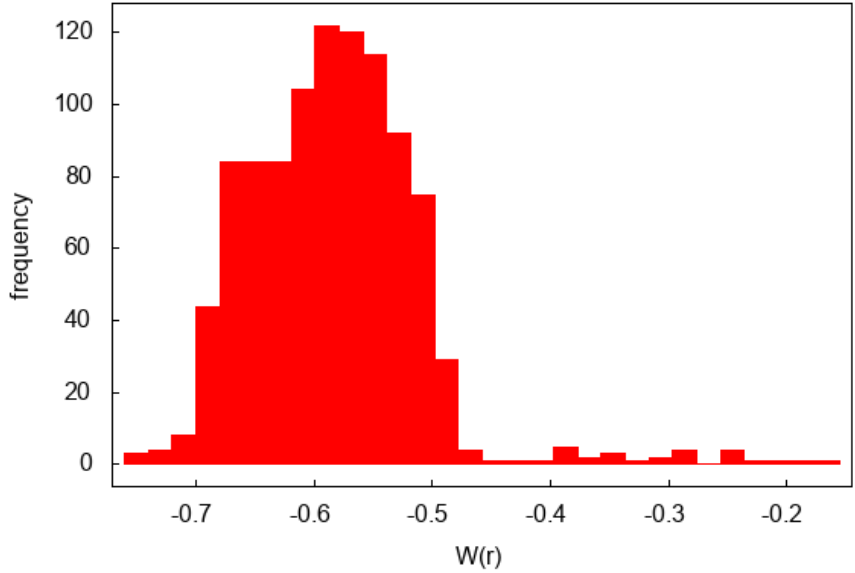}
  \caption{Histogram of $\{W(\vec{r}_{1}),\cdots,W(\vec{r}_{1000})\}$.
  The number of classes is 30.
  Parameters: $q_{C}=0.2$, $q_{I}=0.3$, $q_{O}=0.8$, $N=10$.}
  \label{fig:histogram}
\end{figure}
We wish to demonstrate that $\vec{r}(t)$ converges to the Nash equilibrium point.
When $N(q_{O}-q_{I})\le a+q_{O}$, $\beta(\overline{r}_{n}(t))=0$
such that the right-hand side{} of \eqref{eq:derivative-brd}
is negative whenever $r_{n}(t)>0$.
Therefore, $r_{n}(t)\rightarrow0$ as $t\rightarrow\infty$.
In the $N(q_{O}-q_{I})>a+q_{O}$ case,
we first describe the analytical property of the Nash equilibrium point.
It is asymptotically stable. Indeed, substituting
$r_{n}=r_{\text{Nash}}+\delta r_{n}$ into \eqref{eq:derivative-brd}
yields
\begin{equation*}
  \frac{d}{dt}\delta\vec{r} = A\delta\vec{r},
  \quad
  A =
  \begin{pmatrix}
    \mu & \nu & \cdots & \nu \\
    \nu & \mu & \ddots & \vdots \\
    \vdots & \ddots & \ddots & \nu \\
    \nu & \cdots & \nu & \mu
  \end{pmatrix},
\end{equation*}
where
\begin{equation*}
  \mu = \frac{\partial^{2}w}{\partial r^{2}}(r_{\text{Nash}},r_{\text{Nash}})
  < \nu = \frac{\partial^{2}w}{\partial r\partial\overline{r}}(r_{\text{Nash}},r_{\text{Nash}})
  < 0.
\end{equation*}
The coefficient matrix $A$ contains two eigenvalues,
$\mu+(N-1)\nu$ with algebraic multiplicity 1
and $\mu-\nu$ with algebraic multiplicity $N-1$.
This proves the asymptotic stability
of the Nash equilibrium point.

Next, we numerically demonstrate that an arbitrary solution, $\vec{r}(t)$, of 
the equation \eqref{eq:derivative-brd}
converges to the Nash equilibrium point.
First, we integrate the equation 
\eqref{eq:derivative-brd}
for the $N=2$ case.
Figure \ref{fig:integral-curve}
shows the integral curves of \eqref{eq:derivative-brd} for
various initial points. The equation is symmetrical
with respect to\ the exchange of variables $r_{1}$ and $r_{2}$.
This implies that every solution approaches the Nash equilibrium point
as $t\rightarrow\infty$.

Next, we differentiate the function $V(\vec{x})$, defined in
\eqref{eq:LyapunovFunction}, along a solution of the equation
\eqref{eq:derivative-brd},
\begin{equation*}
  \frac{d}{dt}V(\vec{x}(t))
  = W(\vec{r}(t)),\quad
  W(\vec{r}) = \sum_{n=1}^{N}\frac{\partial w}{\partial r}(r_{n},[\overline{r}_{n}])
  \left\{x_{n} + \sum_{k=1}^{N}x_{k}\right\}.
\end{equation*}
We randomly choose 1000 points, $\vec{r}_{k}\in J$, $k=1,\cdots,1000$
and calculate $W(\vec{r}_{k})$.
Figure \ref{fig:histogram} shows the histogram of data
$\{W(\vec{r}_{k})\}_{k=1,\cdots,1000}$.
This suggests that $W(\vec{r})$ is negative on $J$.

The results above strongly indicate that $\vec{r}(t)$
converges to the Nash equilibrium point.

\subsection{Cooperative Dynamics}
When agents are cooperative, the system may 
converge to the Pareto optimal state.
A possible dynamical system that approaches the Pareto optimal point
can be expressed as follows:
\begin{equation*}
  \frac{dr_{n}}{dt} = \frac{\partial I}{\partial r_{n}}(\vec{r}),\quad
  n=1,\cdots,N,
\end{equation*}
where $I(\vec{r})$ is the total fitness function defined in
\eqref{eq:total_fitness_function}.
When the initial point is close to the Pareto optimal point
$\vec{r}_{\text{Pareto}}$, this system will converge to it because
$\vec{r}_{\text{Pareto}}$ is the asymptotically stable unique fixed point of the system.
Indeed, the Hessian of $I$ at $\vec{r}_{\text{Pareto}}$ is positive definite.
See Appendix.

However, this intuitive Pareto dynamics presents some deficiencies.
First, the solution is not guaranteed to remain in $J$.
Next, it is difficult to know whether the Pareto optimal point is
a global attraction point\footnote{Saddle points exist outside 
the domain $J$.}.

\begin{figure}[h]
  \centering
  \includegraphics[width=16cm]{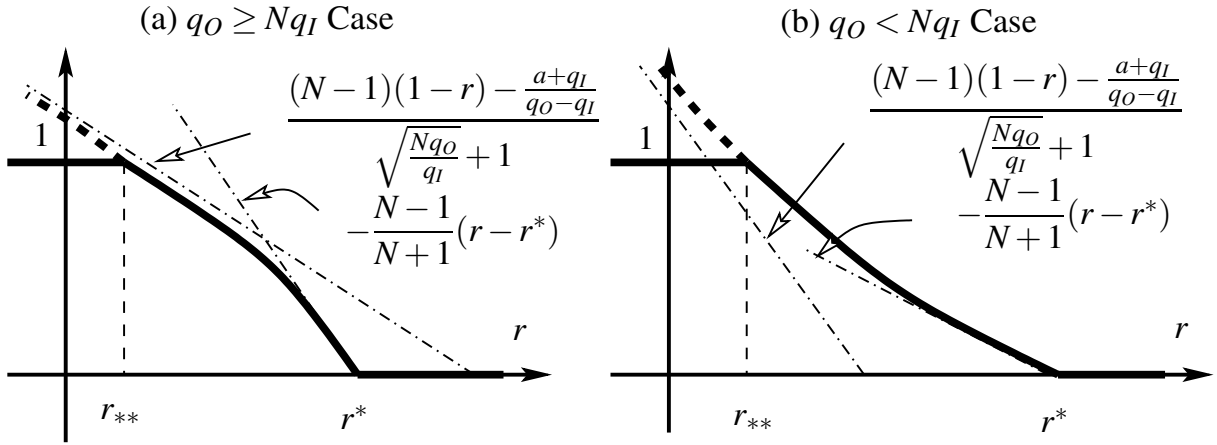}
  \caption{Plots of the function $\beta_{P}(r)$.
  The entity $r_{\ast\ast}$ is given by
  $r_{\ast\ast}
  = 1-\frac{q_{I}(a+q_{O})-a(q_{O}-q_{I})+\sqrt{D_{3}}}{2(N-1)q_{I}(q_{O}-q_{I})}$,
  where $D_{3}=\{q_{I}(a+q_{O})-a(q_{O}-q_{I})\}^{2}
  + 4q_{I}(q_{O}-q_{I})(a+q_{O})(a+Nq_{O})$.
  The entity $r^{\ast}$ is defined in
  fig.~\ref{fig:beta}.
  }
  \label{fig:paretofunc}
\end{figure}

Therefore, we will introduce another dynamical system with better properties,
that is, a Pareto version of the best response dynamics.
A natural definition of the best response function of the Pareto type 
for agent $n$ may be $\displaystyle{\mathop{\mathrm{argmax}}\limits_{0\le r_{n}\le1}I(\vec{r})}$.
This is the larger zero $r_{n}$ of the function,
\begin{equation*}
  \frac{\partial I}{\partial r_{n}}(\vec{r})
  = \frac{\partial w}{\partial r}(r_{n},\overline{r}_{n})
  + \frac{1}{N-1}\sum_{k\neq n}\frac{\partial w}{\partial\overline{r}}(r_{k},\overline{r}_{k}),
\end{equation*}
truncated to the closed interval $[0,1]$.
Unfortunately, the resulting function is not desirable because of its complexity.
Instead, we adopt the following simple definition of the best response
function $\beta_{P}(\overline{r})$ of the Pareto type,
\begin{align}
  \label{eq:betap}
  \beta_{P}(\overline{r}) &= \max(0,\min(1,\overline{\beta}_{P}(\overline{r}))),\quad\text{where}
  \\
  \overline{\beta}_{P}(\overline{r}) &= \text{larger zero $r$ of}\;
  \frac{\partial w}{\partial r}(r,\overline{r}) + \frac{\partial w}{\partial\overline{r}}(r,\overline{r}).
  \notag
\end{align}
This is an extension of $\beta_{N}(\overline{r})$, the best response function
of the Nash type.
The features of the function $\beta_{P}(r)$
are summarized in fig.~\ref{fig:paretofunc}.
Additionally, it exhibits the following properties:
(i) the inequality $\beta_{P}(r)\le\beta_{N}(r)$ is true,
and (ii) $r_{\text{Pareto}}$ defined in \eqref{eq:ParetoOptimal}
is the unique fixed point.

Now, we introduce the cooperative dynamical system, that is,
the best response dynamical system of the Pareto type:
\begin{equation}
  \label{eq:cooperative}
  \frac{dr_{n}}{dt} = \beta_{P}(\overline{r}_{n}) - r_{n},\quad
  n=1,\cdots,N.
\end{equation}
This is well defined because $0\le\beta_{P}\le1$ such that
$\vec{r}(t)\in J$ is guaranteed.
The inequality $\beta_{P}(r)\le\beta_{N}(r)$ suggests that
agents obeying \eqref{eq:cooperative} can be described as
more cooperative than in the case of \eqref{eq:continuous-brd}.

We shall demonstrate that all the solutions of \eqref{eq:cooperative}
converge to the Pareto optimal point $\vec{r}_{\text{Pareto}}$.
Hence, we present the following estimation
derived from fig.~\ref{fig:paretofunc}:
\begin{gather}
  \beta_{P}(r) - \beta_{P}(r_{\text{Pareto}})
  = -(N-1)\gamma_{P}(r)(r-r_{\text{Pareto}}),
  \label{eq:betap-estimation}
  \\
  0\le\gamma_{P}(r)
  \le \max\left(\frac{1}{N+1}, \frac{1}{\sqrt{\frac{Nq_{O}}{q_{I}}}+1}\right)
  \equiv c
  < 1.
  \notag
\end{gather}
This is an analogy of \eqref{eq:beta-estimation}.
Let $x_{n}=r_{n}-r_{\text{Pareto}}$ and let $\gamma_{n}(t)=\gamma_{P}(\overline{r}_{n}(t))$.
Subsequently, we have a relation analogous to \eqref{eq:beta-gamma}:
\begin{equation}
  \label{eq:betap-gammap}
  \beta_{P}(\overline{r}_{n}(t)) - r_{n}(t)
  = -\gamma_{n}(t)\sum_{k\neq n}x_{k}(t) - x_{n}(t).
\end{equation}
Further, the time derivative of the function $V(\vec{x})$
defined in \eqref{eq:LyapunovFunction}
along the solution of \eqref{eq:cooperative} reads
\begin{equation*}
  \frac{d}{dt}V(\vec{x}(t))
  \le -2(1-c)V(\vec{x}(t)),
  \quad\therefore
  0 \le V(\vec{x}(t)) \le V(\vec{x}(0))e^{-2(1-c)t}
  \rightarrow0\quad(t\rightarrow\infty).
\end{equation*}
Therefore, we conclude that all solutions of \eqref{eq:cooperative}
converge to the Pareto optimal point $\vec{r}_{\text{Pareto}}$.

It is noteworthy that a discrete version of the system \eqref{eq:cooperative},
\begin{equation}
  \label{eq:discrete-Pbrd}
  r_{n}(t+1) = r_{n}(t) + \alpha\{\beta_{P}(\overline{r}_{n}(t)) - r_{n}(t)\},\quad
  n=1,\cdots,N,
\end{equation}
converges to the Pareto optimal point as well if the learning rate $\alpha$
is sufficiently small.

Equation \eqref{eq:discrete-Pbrd} is a model equation of
the simultaneous-update type (see also eq.~\eqref{eq:discrete-brd}).
An individual update-type model is realized by
replacing the learning rate $\alpha$ with $\alpha/N$
in eq.~\eqref{eq:discrete-Pbrd}.

\section{Experimental Studies}
\label{sec:Experiments}
We have performed an experiment to study whether human adopts the Pareto
equilibrium.
It is a laboratory experiment performed in Kitasato and
Hirosaki University.
In the experiment, multiple human players participate in
the game and compete for the number of coins to be acquired.
A total of thirty three  subjects (1 female and 32 males; mean age(1 s.d.)
=20.2(1.6)) participated and we label them as $i\in \{1,\cdots,33\}$.
There are 11 groups of three subjects ($N=3$) and we label them as
$G\in \{1,\cdots,11\}$. The subjects in each group know each other and
the total reward given to the participants is in proportion to the total
number of coins acquired by them (3 yen/coin).
  It is an incentive to cooperate with each other in the same group
  to maximize the number of coins.
The optimal strategy of the players is $r_{\text{Pareto}}$.
The reward is in the range of 2000 yen and 3000 yen per subject.

  There are three rounds of thirty minutes and we label
  them as $R\in \{1,2,3\}$. Between the rounds, there are two
  intervals of 10 minutes and the players can
  discuss how to maximize the number of  coins.
  The optimal strategy is $r_{\text{Pareto}}$.
  We adopt the parameter setting $q_{I}=0.1,q_{O}=0.8,q_{C}=0.2$
  so that $r_{\text{Pareto}}=0.23$ and $r_{\text{Nash}}=0.34$.
Table \ref{exp:tab1} summarizes the settings of the experiment.

\begin{figure}[htbp]
\begin{center}
\begin{tabular}{cc}
\includegraphics[width=6cm]{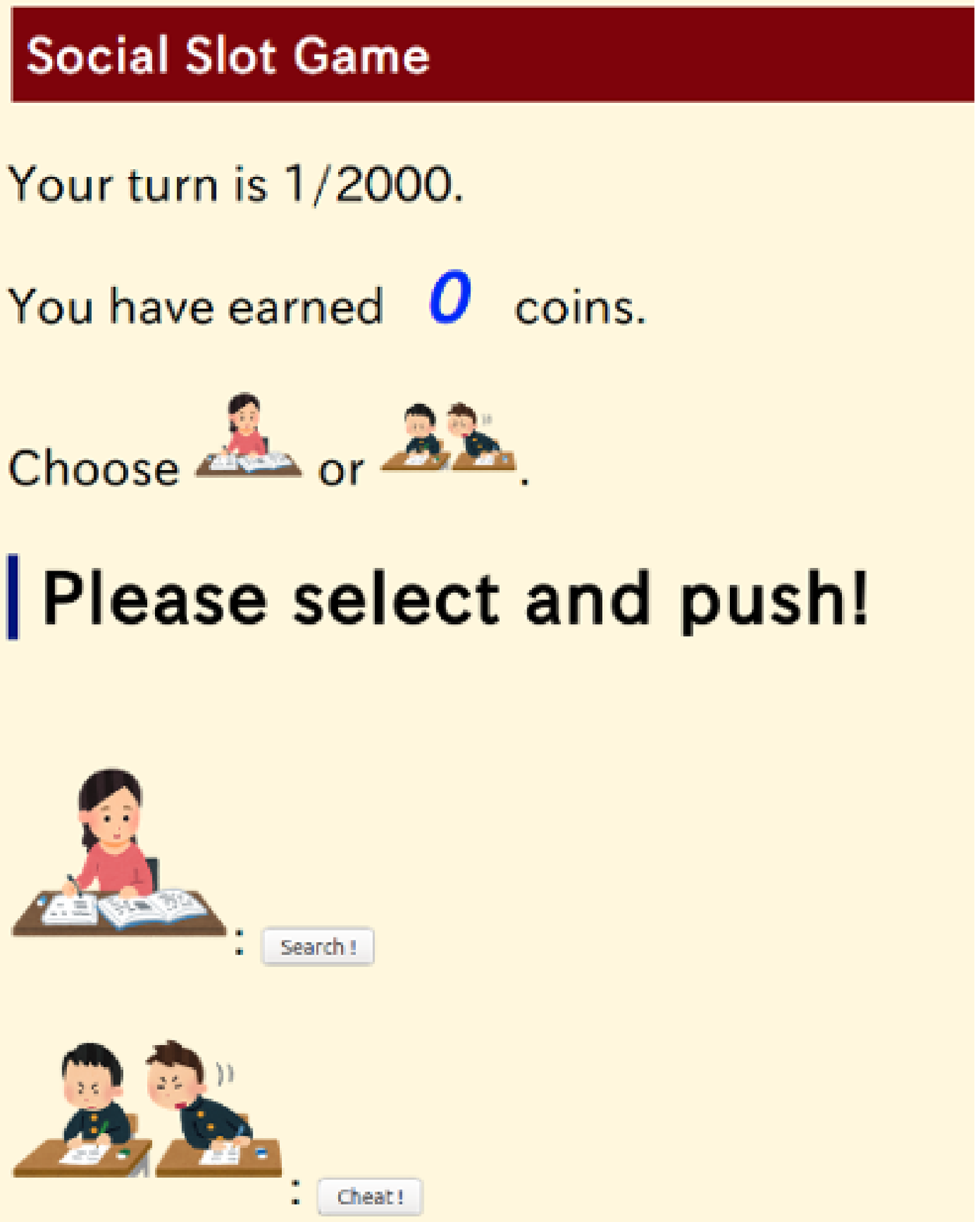}
&
\includegraphics[width=6cm]{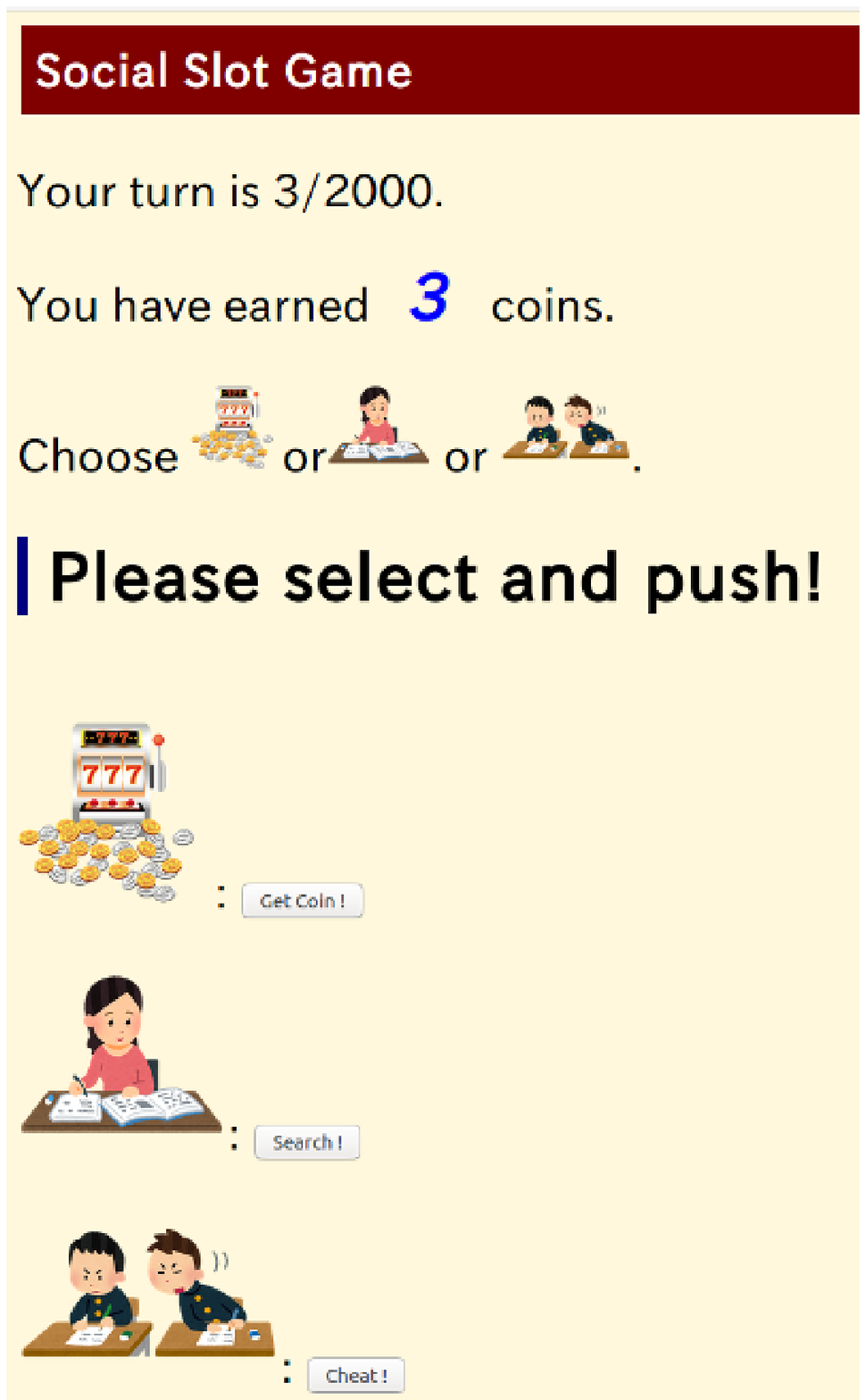} 
\end{tabular}
\end{center}
\caption{Screenshot of the game. Left figure
  shows the screen when the player does not know the bandit
  with $\sigma=1$. Individual learning and social learning are
  chosen by pushing the button, "Search ! " and "Cheat !", respectively.
  If he knows the bandit with $\sigma=1$, he sees the screen in
  the right figure. Usually, he exploits the bandit by pushing the button
  "Get Coin !". 
}
\label{exp:fig1}
\end{figure}

\begin{table*}[htbp]
\begin{center}
\begin{tabular}{ccccccc}
\hline
$N$ &  $S$ & $T$ & $R$ & Reward & Date & Subject Pool \\ 
3 & 33 & 263 & 1 & 3 yen/coin &2017/6 and 2017/11 & Both Univ.   \\
3 & 33 & 319 & 2 & 3 yen/coin &2017/6 and 2017/11 & Both Univ.   \\
3 & 33 & 322 & 3 & 3 yen/coin &2017/6 and 2017/11 & Both Univ.   \\
\hline
\end{tabular}
\caption{\label{exp:tab1}
  Experimental design. $N$: the number of subjects in each group,
  $S$: the total number of subjects, $T$: average number
  of turns (actions) of the subjects, $R$: the round number.
}
\end{center}
\end{table*}

\subsection{Method}

We have developed a browse rMAB game. The game interface is shown in
Figure \ref{exp:fig1}.
The experiments were performed in 
the laboratory. Experimenter explained the experimental procedures and 
the rewards, the subjects were asked to sit on chairs that are located 
far from each others. While playing the game, it was forbidden to talk 
to other subjects so that it was impossible to share the information whether
they know the bandit with $\sigma=1$ or not.
During the intervals, the subjects could freely
talk with others and discussed how to get more coins.

Figure \ref{exp:fig1} shows the screenshot of the game.
When the player does not know the bandit with $\sigma=1$,
he sees the left figure.
There are two options, individual learning and social learning,
which are chosen by pushing the buttons with
labels, "search !" and "cheat !", respectively.
When the player knows the bandit with $\sigma=1$, he sees the right figure.
Usually, the player exploits the bandit by pushing the button with label
"Get Coin !" and get some coins.
The number of coins is in the range $\{1,2,3\}$ and it
is fixed at random with probability 1/3
when one finds the bandit.
This mechanism was introduced to make the game interesting.
The other two buttons provide the player to
changes the bandit. If the number of coins
in the exploit is one, one should want to changes the bandit.
If one push "Search" or "Cheat !", one searches
the bandit with $\sigma=1$ again. If he succeeds, he can change the
number of the coins. "Cheat" button is better than
"Search" button when $q_{O}>q_{I}$. The subjects were taught about
 the function before the start of the experiment.
Hereafter, we explain the results of the experiment.
Summary of the statistics are given in Table \ref{exp:tab2}.

\subsection{Results}

\begin{figure}[htbp]
\begin{center}
\begin{tabular}{cc}
\includegraphics[width=8cm]{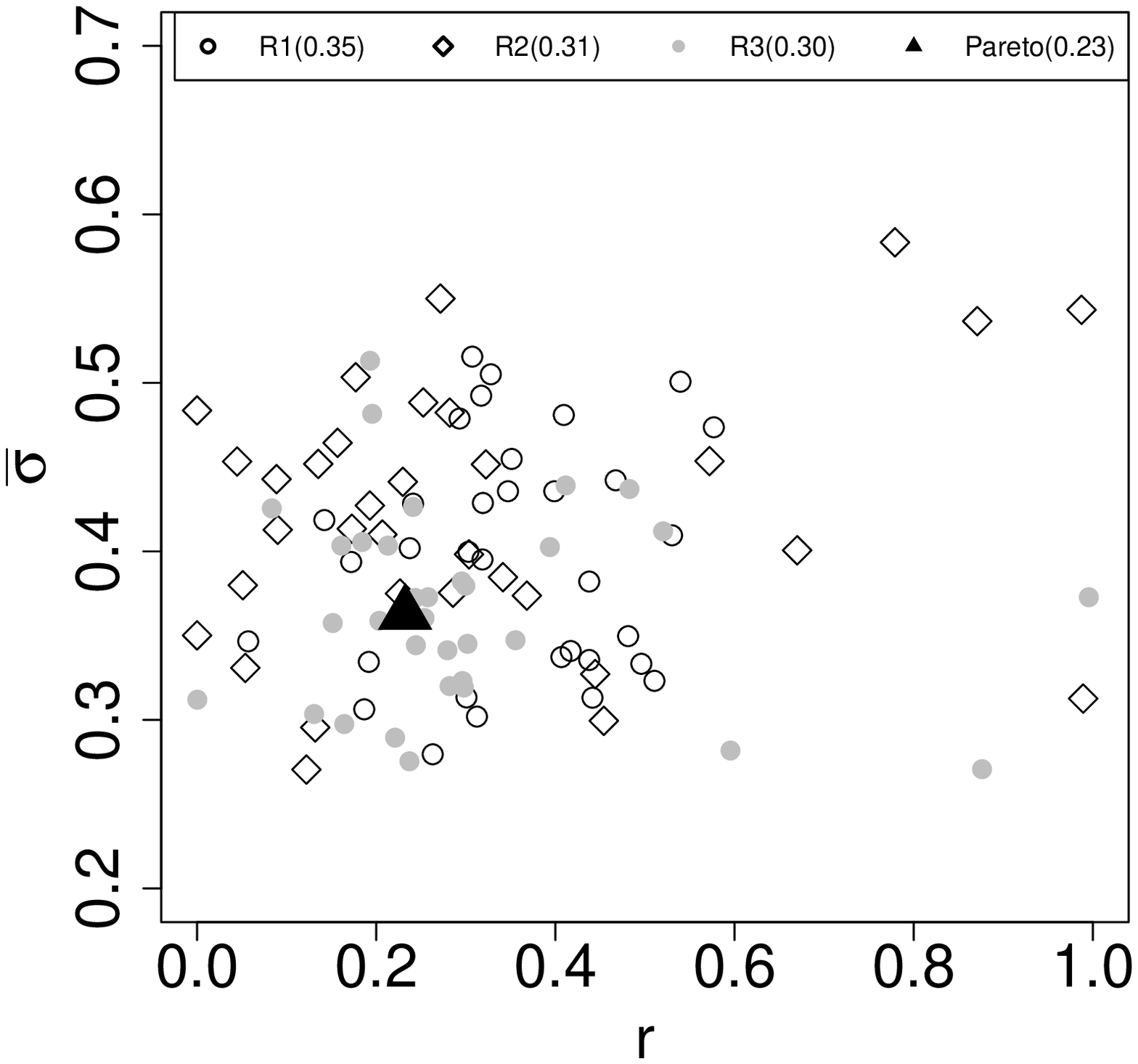}
&
\includegraphics[width=8cm]{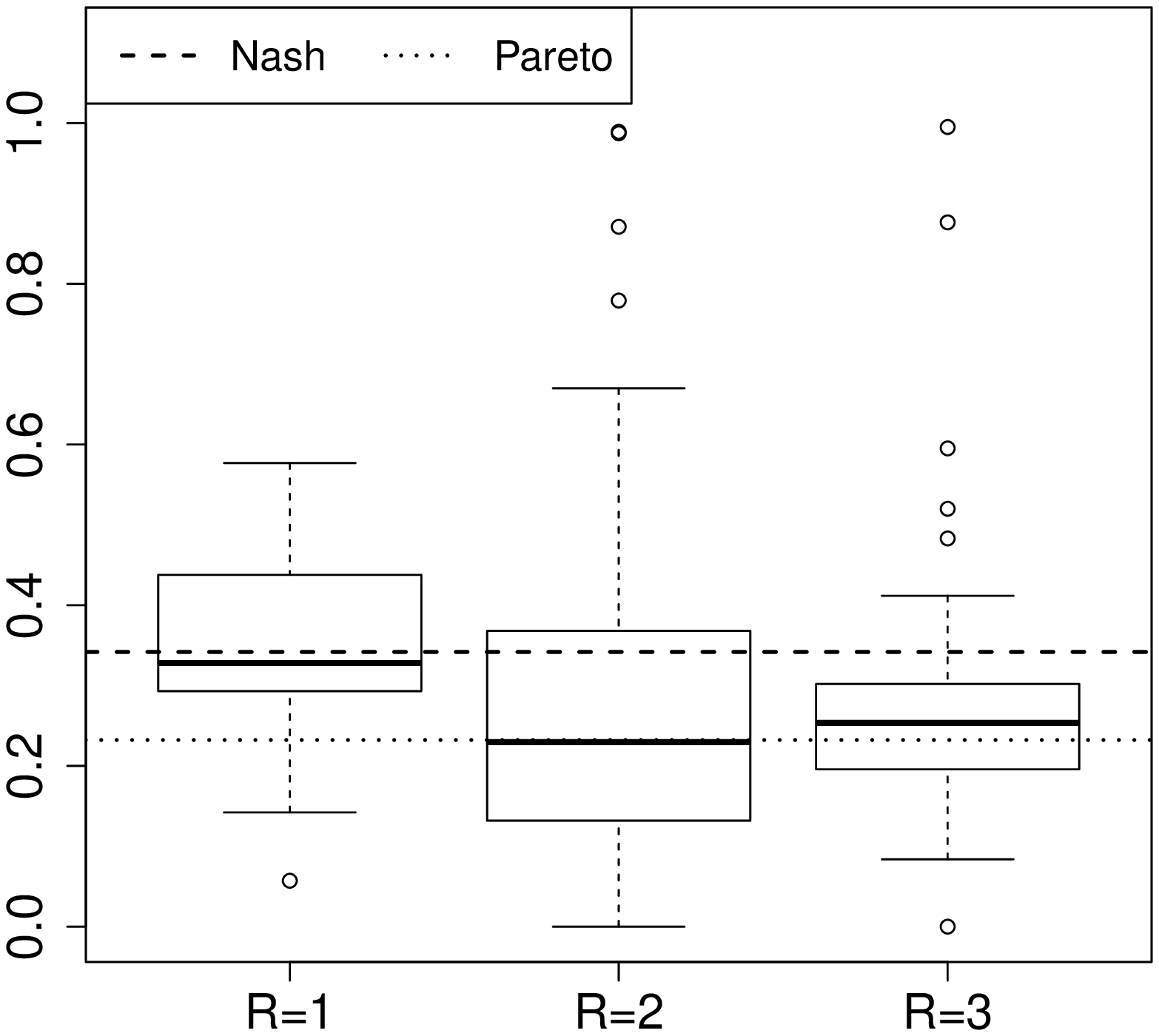} 
\end{tabular}
\end{center}
\caption{Left: 
  Scatter plot of $(r_i,\overline{\sigma}_i)$. There are three rounds in the experiment, $R\in \{1,2,3\}$.
  Empty circles, empty diamonds and gray circles  show
  the plot for $R=1,2,3$, respectively.
  The black triangle shows $(r_{\text{Pareto}},w_{\text{Pareto}})$.
  Right: Boxplot of $r_{i}$.
  The broken (dotted) horizontal line shows $r_{\text{Nash}}$ ($r_{\text{Pareto}}$). 
}
\label{exp:fig4}
\end{figure}

The left figure of Figure \ref{exp:fig4} shows
the scatter plot of $(r_i,\overline{\sigma}_i)$.
There are three rounds in the experiment, $R\in \{1,2,3\}$,
we show the plots with different symbols for the rounds.
The black triangle shows $(r_{\text{Pareto}},w_{\text{Pareto}})$.
The right figure shows the boxplots of $r_i$.
The broken and dotted lines shows $r_{\text{Nash}}$ and $r_{\text{Pareto}}$.
The collective average of $r_i$ in round 2 and round 3
are 0.31 and 0.30 and they are slightly larger than $r_{\text{Pareto}}=0.23$.
The median (50 percentile) of $r_i$ in round 2 and round 3 are 0.23 and 0.25,
which almost coincide with $r_{\text{Pareto}}$. As the experiment proceeds,
we observe more collaborative behavior.

\begin{table*}[htbp]
\begin{center}  
\begin{tabular}{lcccccc}
\hline
$R$ & $r_{opt}$ &  Mean of $r_{i}(SE)$ & SD of $r_i$ & Median of $r_{i}$ & $w(r_{opt},r_{opt})$
& Mean of $\overline{\sigma}_i(SE)$  \\ 
\hline
1 & 0.23 & 0.35(2) &0.07 & 0.33 & 0.36& 0.40(1) \\
2 & 0.23 & 0.31(5) &0.08 & 0.23 & 0.36& 0.42(1) \\ 
3 & 0.23 & 0.30(4) &0.06 & 0.25 & 0.36& 0.37(1) \\
\hline
\end{tabular}
\caption{\label{exp:tab2}
  Summary of experimental results.
$r_{opt}$ is the optimal strategy.}
\end{center}
\end{table*}

In order to test the above hypothesis, we
adopt a hierarchical Bayesian method.
We assume $r_i$ for subject $i$
who participate in the experiment of  group $G[i]\in \{1,\cdots,11\}$
in round $R[i]\in \{1,2,3\}$  obeys,
\begin{eqnarray}  
 r_i &\sim & \mbox{t}(\nu=3,\mu=r(R[i],G[i]),\mbox{scale}=\sigma_{A}) \nonumber \\
 r[R,G]&=&r_{A}+\Delta r_R[R]+\Delta r_G[G] \nonumber \\
 \Delta r_R[R]&\sim& \mbox{N}(0,\sigma_R^2) \nonumber \\
 \Delta r_G[G]&\sim& \mbox{N}(0,\sigma_{G}^2) \nonumber 
\end{eqnarray}
$\Delta r_R$ and $\Delta r_G$ describe the dependence
of $r(R,G)$ on $R$ and $G$, respectively.
As for $\sigma_{A},\sigma_R,\sigma_G$, we assume
the Half-Cauchy prior,
\[
\sigma_{A},\sigma_{R},\sigma_{G}\sim \mbox{Cauchy}_{+}(25).
\]
We studied the posterior distribution of $r_{A},\Delta r_R[R],\Delta
r_G[G]$ and estimate the 95\% Bayesian credible intervals using Stan 2.19.2
in R 3.6.2 software.
We have checked the convergence of the sampling by the Gelman-Rubin statistics.
Table \ref{exp:tab3} shows the results for $r_{A}+\Delta r_R[R]$.

\begin{table*}[htbp]
\begin{center}  
\begin{tabular}{cccc}
\hline
 & $R=1$ & $R=2$ & $R=3$ \\ 
\hline
mean & 0.34  & 0.24  & 0.26   \\
95\% C.I. & [0.28,0.41]  & [0.17,0.30]  & [0.20,0.31] \\
\hline
\end{tabular}
\caption{\label{exp:tab3}
The mean and 95\% Bayesian credible intervals of $r_{A}+\Delta r_{R}[R]$.    
}
\end{center}
\end{table*}

We see that the credible intervals of $r_{A}+\Delta r_{R}[R]$
for $R=2,3$ does not include $r_{\text{Nash}}=0.34$.
We also reject the hypothesis that $\Delta r_R[R=1]=\Delta r_R[R=2]$
,$\Delta r_{R}[R=1]=\Delta r_{R}[R=3]$ with significance of 1\%
by estimating 
$P(\Delta_{R}[1]<\Delta_{R}[2])$ and $P(\Delta_{R}[1]<\Delta_{R}[3])$.
We observe collaborative behaviors in $R=2$ and $R=3$.

\section{Concluding Remarks}
We herein introduced a multiagent system in a restless 
multiarmed bandit game and studied the optimal learning dynamics of agents
theoretically and experimentally.

As is well known, the Nash equilibrium is a typical solution in
noncooperative games. We demonstrated that best response dynamics
drove the system to equilibrium.
In the case of cooperative game,
we introduced a new dynamical system, that is, a best response dynamical
system of the Pareto type and proved that it converged to the Pareto optimal point.
We also conducted a cooperative game type experiment.
As shown in fig.\ref{exp:fig4},
the distribution of strategies of the participants appeared
to be centered around the Pareto optimal point as the round proceeded.
This observation was supported by a hierarchical Bayesian analysis.

The following issues can be explored in the future.
First, we have introduced a new concept of the best response function
in section~\ref{sec:OptimalStrategy},
that is, the best response function of the Pareto type.
It is a natural extension of the function of Nash type.
It may be possible to define the same type of function in generic systems, and
we believe that it is a useful concept.
Second, it would be interesting to perform noncooperative game type experiments
to examine whether humans adopt the Nash strategy.
Third, we investigated the agents' behavior based on the optimal strategies
in equilibrium.  This procedure is reasonable because,
as shown in section~\ref{sec:OptimalStrategy}, the Nash equilibrium
and Pareto optimal states are the natural destinations in the long run.
The experiments on Nash equilibrium
may be explained better if a time-dependent theory is considered.

\section*{Acknowledgments}

This work was supported by JPSJ KAKENHI (Grant No. 17K00347).

\section*{Competing Interests}

The authors declare that they have no competing interests.

\section*{Ethics Declarations}

Informed consent was obtained from all individual participants
involved in the study.

\bibliographystyle{elsarticle-num}

\newpage

\appendix

\renewcommand{\appendixname}{Appendix }

\section*{Appendix\quad Maximum Point of the Total Fitness Function}

Derivative of the total fitness function, $I(\vec{r})$, is given by
\begin{align}
  \frac{1}{a}\frac{\partial I}{\partial r_{n}}(\vec{r})
  &= \frac{1}{a+\kappa}\left\{
    \frac{\kappa(q_{O}-q_{I})-q_{I}(a+q_{O})}{(a+q_{n})^{2}}
    - K\right\},
  \label{eq:dI}
  \\
  K &= \sum_{k=1}^{N}\frac{q_{I}q_{O}r_{k}}{(a+q_{k})(a+\kappa)},\quad
  q_{n} \equiv q_{I} + (q_{O}-q_{I})r_{n}.
  \notag
\end{align}
Because $J$ is compact,
at least one maximum point of $I(\vec{r})$ in $J$ exists.

\paragraph{Maximum Point in $J^{\circ}$}

Let $\vec{r}$ be one of the maximum points of $I(\vec{r})$.
First, we assume that $\vec{r}$ is an inner point of $J$: $\vec{r}\in J^{\circ}$.
This is a zero of \eqref{eq:dI}. 
Therefore, we have
\begin{gather*}
  (q_{O}-q_{I})\kappa - q_{I}(a+q_{O}) > 0,
  \\
  a+q_{n} = \sqrt{\frac{(q_{O}-q_{I})\kappa - q_{I}(a+q_{O})}{K}}.
\end{gather*}
Thus, this point is on the diagonal:
$r_{1}=\cdots=r_{N}=r>0$.
Therefore, this point is also a maximum point of $w(r,r)$.
This is merely the unique maximal point, $r_{\text{Pareto}}$,
of $w(r,r)$. Therefore, $\vec{r}$ is the unique maximum point of
$I(\vec{r})$ in $J^{\circ}$.
It is straightforward to verify that
\begin{equation*}
  (q_{O}-q_{I})\kappa - q_{I}(a+q_{O}) > 0
  \iff N(q_{O}-q_{I}) - (a+q_{O})>0.
\end{equation*}

\paragraph{Maximum Point on $J^{b}$}

Next, we assume that the maximum point, $\vec{r}$, is a boundary point of $J$: $\vec{r}\in J^{b}$.

First, we demonstrate that ${}^{\forall}r_{n}$ equals $0$ or $1$.
Let ${}^{\exists}r_{j}\in(0,1)$.
Subsequently
\begin{equation*}
  \frac{1}{a}\frac{\partial I}{\partial r_{j}}
  = \frac{1}{a+\kappa}\left\{
    \frac{(q_{O}-q_{I})\kappa - q_{I}(a+q_{O})}{(a+q_{j})^{2}}
    - K\right\} = 0.
\end{equation*}
Because $\vec{r}$ is a boundary point, $r_{n}$ equals $0$ or $1$.
However, when $r_{n}=0$, we have
\begin{equation*}
  \frac{1}{a}\frac{\partial I}{\partial r_{n}}
  = \frac{1}{a+\kappa}\left\{
    \frac{(q_{O}-q_{I})\kappa - q_{I}(a+q_{O})}{(a+q_{n})^{2}}
    - K\right\} > 0.
\end{equation*}
When $r_{n}=1$, we have
\begin{equation*}
  \frac{1}{a}\frac{\partial I}{\partial r_{n}}
  = \frac{1}{a+\kappa}\left\{
    \frac{(q_{O}-q_{I})\kappa - q_{I}(a+q_{O})}{(a+q_{n})^{2}}
    - K\right\} < 0.
\end{equation*}
This contradicts the maximality of $\vec{r}$.

Next, we demonstrate that ${}^{\forall}r_{n}$ equals $0$.
Let $r_{1}=\cdots=r_{k}=0$, $r_{k+1}=\cdots=r_{N}=1$.
Subsequently, $\kappa=kq_{I}$ and
\begin{equation*}
  \frac{1}{a}\frac{\partial I}{\partial r_{n}}
  = \frac{1}{a+kq_{I}}\left\{
    \frac{(q_{O}-q_{I})lq_{I}-q_{I}(a+q_{O})}{(a+q_{n})^{2}}
    - K\right\},\quad
  K = \frac{(N-k)q_{I}q_{O}}{(a+q_{O})(a+kq_{I})}.
\end{equation*}
Because $\vec{r}$ is a maximal point of $I(\vec{r})$, we have
\begin{align}
  \frac{1}{a}\frac{\partial I}{\partial r_{1}}
  = \frac{1}{a+kq_{I}}\left\{
    \frac{(q_{O}-q_{I})kq_{I}-q_{I}(a+q_{O})}{(a+q_{I})^{2}}
    - K\right\} \le 0,
  \label{eq:boundary1}
  \\
  \frac{1}{a}\frac{\partial I}{\partial r_{N}}
  = \frac{1}{a+kq_{I}}\left\{
    \frac{(q_{O}-q_{I})kq_{I}-q_{I}(a+q_{O})}{(a+q_{O})^{2}}
    - K\right\} \ge 0.
  \label{eq:boundaryN}
\end{align}
In the case of $k=0$, \eqref{eq:boundaryN} reads
\begin{equation*}
  q_{I}\frac{-(a+q_{O})}{(a+q_{O})^{2}a}
  - \frac{Nq_{I}q_{O}}{(a+q_{O})a^{2}} \ge 0.
\end{equation*}
This cannot occur.
Next, we consider the $1\le k\le N-1$ case.
From \eqref{eq:boundary1} and \eqref{eq:boundaryN},
we have
\begin{equation*}
  q_{I}\frac{(q_{O}-q_{I})k - (a+q_{O})}{(a+q_{I})^{2}(a+kq_{I})}
  \le \frac{(N-k)q_{I}q_{O}}{(a+q_{O})(a+kq_{I})^{2}}
  \le q_{I}\frac{(q_{O}-q_{I})k - (a+q_{O})}{(a+q_{O})^{2}(a+kq_{I})}.
\end{equation*}
This is also impossible because $0<q_{I}<q_{O}$.
Finally, we investigate the $k=N$ case.
The inequality \eqref{eq:boundary1} reduces to
\begin{equation*}
  (q_{O}-q_{I})N - (a+q_{O}) \le 0.
\end{equation*}
This is the only possible case.
That is, if the maximum point of $I(\vec{r})$ is on the boundary $J^{b}$,
it is merely the origin $\vec{r}=\vec{0}$.
In this case, the origin is indeed the maximum point of $I(\vec{r})$:
because
\begin{equation*}
  (q_{O}-q_{I})\kappa - q_{I}(a+q_{O})
  = q_{I}\{(q_{O}-q_{I})N - (a+q_{O})\} \le 0,
\end{equation*}
\eqref{eq:dI} is negative on $J$ except for the origin.

\paragraph{Strict Maximality}

We explicitly demonstrate that $r_{\text{Pareto}}$ is strictly maximal when
$r_{\text{Pareto}}\in J^{\circ}$.
We consider the Hessian of the total fitness function at
the Pareto optimal point,
\begin{align*}
  H(\vec{r}_{\text{Pareto}}) &= \frac{1}{2}\left(\frac{\partial^{2}I}{\partial r_{m}\partial r_{n}}(\vec{r}_{\text{Pareto}})\right)
  = -\begin{pmatrix}
  \alpha+\beta & \beta & \cdots & \beta \\
  \beta & \alpha+\beta & \ddots & \vdots \\
  \vdots & \ddots & \ddots & \beta \\
  \beta & \cdots & \beta & \alpha+\beta
  \end{pmatrix},
\end{align*}
where
\begin{align*}
  \alpha &= a\frac{(q_{O}-q_{I})\{(q_{O}-q_{I})\kappa_{\text{Pareto}} - q_{I}(a+q_{O})\}}{(a+q_{\text{Pareto}})^{3}(a+\kappa_{\text{Pareto}})}
  \\
  \beta &= \frac{aq_{I}q_{O}(a+qI_{I})}{(a+q_{\text{Pareto}})^{2}(a+\kappa_{\text{Pareto}})^{2}}
  + \frac{Naq_{I}^{2}q_{O}r_{\text{Pareto}}}{(a+q_{\text{Pareto}})(a+\kappa_{\text{Pareto}})^{3}},
\end{align*}
and
\begin{equation*}
  \kappa_{\text{Pareto}} = Nq_{I}(1-r_{\text{Pareto}}),\quad
  q_{\text{Pareto}} = q_{I}+(q_{O}-q_{I})r_{\text{Pareto}}.
\end{equation*}
The Hessian contains two eigenvalues,
$-\alpha$ with algebraic multiplicity $(N-1)$,
and $-(\alpha+N\beta)$ with algebraic multiplicity $1$.
Because
\begin{equation*}
  (q_{O}-q_{I})N-(a+q_{O}) > 0 \iff
  (q_{O}-q_{I})\kappa_{\text{Pareto}} - q_{I}(a+q_{O}) > 0,
\end{equation*}
we obtain $\alpha,\beta>0$.
This proves the strict maximality.

\end{document}